\documentclass[letterpaper, 10 pt, conference]{ieeeconf} 

\IEEEoverridecommandlockouts                              

\renewcommand{\baselinestretch}{0.955}

\usepackage{amsmath}
\usepackage{amssymb}
\usepackage{bm}
\usepackage{mathrsfs}
\usepackage{tikz}

\usepackage{amsthm}
\usepackage{xspace}
\usepackage{array}
\usepackage{multirow}
\usepackage{cite}
\usepackage{lipsum}

\usepackage{colortbl}
\definecolor{cello}{HTML}{ffe6cc}

\usepackage{graphics} 

\usepackage{color}
\usepackage{xcolor}
\usepackage{accents}
\usepackage{centernot}
\usepackage{algorithm}
\usepackage[noend]{algorithmic}
\usepackage[font=small, labelfont=bf]{caption}
\usepackage{mathtools}
\usepackage{hyperref}
\usepackage{lipsum}

\newcommand{\dimst}{{n_x}}
\newcommand{\ssp}{\mathcal{X}}
\newcommand{\compacts}{\mathbb{K}\ssp}
\newcommand{\ti}{t}
\newcommand{\tf}{T}
\newcommand{\tint}{[\ti,\tf]}
\newcommand{\cost}{C}
\newcommand{\tj}{\xi_f}

\newcommand{\tgfn}{J}
\newcommand{\tg}{\mathcal{T}}
\newcommand{\cset}{\mathcal{U}}
\newcommand{\dset}{\mathcal{D}}
\newcommand{\csetsig}{\mathbb{U} (\ti)}
\newcommand{\dsetsig}{\mathbb{D} (\ti)}
\newcommand{\strd}{\mathfrak{d}}
\newcommand{\stratset}{\mathfrak{D}(t)}
\newcommand{\stre}{\mathfrak{e}}
\newcommand{\strateset}{\mathfrak{E}(t)}
\newcommand{\hjrset}{\mathcal{R}}
\newcommand{\fseto}{\hat{\fset}}
\newcommand{\fset}{\mathcal{S}}
\newcommand{\ftube}{\bar{\fset}}

\newcommand{\ham}{H}
\newcommand{\maxou}{\max_{u \in \cset}}
\newcommand{\minou}{\min_{u \in \cset}}
\newcommand{\maxod}{\max_{d \in \dset}}
\newcommand{\minod}{\min_{d \in \dset}}

\newcommand{\dimsa}{{n_k}}
\newcommand{\sasp}{\mathcal{G}}
\newcommand{\lf}{\Psi}
\newcommand{\pj}{P}
\newcommand{\rg}{\mathscr{R}}
\newcommand{\rglf}{{\rg}_\lf} 
\newcommand{\rglftg}{{\rg}_{\lf | _\tg}}
\newcommand{\rglffeas}{{\rg}_{\lf | _{\ftube(\tg, t)}}}

\newcommand{\lfl}{\psi}
\newcommand{\lfi}{{\lf}^{-1}}
\newcommand{\fg}{f_\sasp}
\newcommand{\tgg}{\tg_\sasp}

\newcommand{\tgfg}{\tgfn_\sasp}
\newcommand{\Vg}{V_{\sasp}}
\newcommand{\Vge}{V_{\sasp, \maxerr}}
\newcommand{\tjg}{\xi_{\fg}}

\newcommand{\maxerr}{\delta^*}
\newcommand{\eps}{\varepsilon}
\newcommand{\errset}{\mathcal{E}}
\newcommand{\errsigset}{\mathbb{E} (\ti)}
\newcommand{\fglin}{\kappa}
\newcommand{\tjgk}{\xi_{\fglin + \varepsilon}}

\newcommand{\hamlerr}{\ham_{\maxerr}}

\newcommand{\inn}{\cdot}
\newcommand{\out}{\circ}

\newtheorem{Theorem}{Theorem}
\newtheorem{Lemma}{Lemma}

\newtheorem{Corollary}{Corollary}

\newtheorem{Remark}{Remark}

\makeatletter
\let\NAT@parse\undefined
\makeatother

\title{
State-Augmented Linear Games with Antagonistic Error for High-Dimensional, Nonlinear Hamilton-Jacobi Reachability
}

\author{Will Sharpless, Yat Tin Chow and Sylvia Herbert
\thanks{Sharpless and Herbert are with Univ. of California, San Diego.  
\{\href{mailto:wsharpless@ucsd.edu}{wsharpless}, \href{mailto:sherbert@ucsd.edu}{sherbert}\}@ucsd.edu
This work is supported by ONR YIP N000142212292, ONR grant N000142412661, NSF DMS-2409903 and NIH/HHMI grant T32 EB009380 (McCulloch). The content is the responsibility of the authors.}
}

\begin{document}
\maketitle
\thispagestyle{empty}
\pagestyle{empty}
\begin{abstract}
Hamilton-Jacobi Reachability (HJR) is a popular method for analyzing the liveness and safety of a dynamical system with bounded control and disturbance. The corresponding HJ value function offers a robust controller and characterizes the reachable sets, but is traditionally solved with Dynamic Programming (DP) and limited to systems of dimension less than six. Recently, the space-parallelizeable, generalized Hopf formula has been shown to also solve the HJ value with a nearly three-log increase in dimension limit, but is limited to linear systems. To extend this potential, we demonstrate how state-augmented (SA) spaces, which are well-known for their improved linearization accuracy, may be used to solve tighter, conservative approximations of the value function with any linear model in this SA space. Namely, we show that with a representation of the true dynamics in the SA space, a series of inequalities confirms that the value of a SA linear game with \textit{antagonistic error} is a conservative envelope of the true value function. It follows that if the optimal controller for the HJ SA linear game with error may succeed, it will also succeed in the true system. Unlike previous methods, this result offers the ability to safely approximate reachable sets and their corresponding controllers with the Hopf formula in a non-convex manner. Finally, we demonstrate this in the slow manifold system for clarity, and in the controlled Van der Pol system with different lifting functions.
\end{abstract}

\section{Preamble}

Verifying that a system satisfies safety or goal-satisfaction specifications for nonlinear systems with bounded control and disturbance inputs is a crucial yet computationally challenging task. Hamilton-Jacobi reachability (HJR) analysis is a formal verification tool for guaranteeing the performance and safety of such systems. HJR first defines a cost function that encodes a target set of states to either reach or avoid, and then solves a differential game backward in time between the control and disturbance inputs, assuming the latter is adversarial. The result is a value function that encodes the backward reachable set (BRS): the set of states from which the controller may drive the system into the target despite any disturbance (for the \textit{Reach} problem), or where the optimal control will fail to ultimately avoid the target for the worst-case disturbance (for the \textit{Avoid} problem). Moreover, the corresponding optimal control policy in either case may be solved from the gradient of the value function.

HJR has been widely applied in numerous safety-critical applications \cite{mitchell2005time, bansal2017hamilton} due to its ability to produce strong guarantees. However, HJR relies on dynamic programming (DP), which suffers from exponential compute burden with respect to dimension (i.e., the \textit{curse of dimensionality}). In practice, DP is unable to solve HJ problems with systems of greater than three dimensions online and six offline. Several works have sought to improve scalability via learning methods \cite{bansal2017hamilton, lin2023generating}, linearization \cite{Kirchner_2018, lee19iterativehopf}, and decomposition \cite{chen2017exact, he2023efficient}, but scalability remains a challenge when deterministic guarantees are required.

Solving differential games with linear dynamics and bounded inputs is more tractable due to the recent application of the generalized Hopf formula \cite{darbon2016algorithms, chow2017algorithm}. If the target is convex, by a change of coordinates with the fundamental matrix, the value function may be solved independently in space by optimization over the space of the costate. This allows the game value (and optimal control) to be rapidly solved for a single point in space and time without the exponential burden, and experimentally, has allowed computation of systems of up to dimension 4096 to be solved in milliseconds \cite{chow2019algorithm}. The major limitation is that the system must be linear for the validity of the generalized Hopf formula.

To apply the Hopf formula to nonlinear systems, standard linearizations have been used successfully in multi-agent pursuit-evasion games \cite{Kirchner_2018}, in an iterative, LQR-like fashion \cite{lee19iterativehopf}, and by approximating the Koopman operator \cite{sharpless2023koopman}. Despite empirical success, none of these approaches provides conservative guarantee on the value.
Recently, it was shown that a conservative solution may be derived by transforming the error between any linear model and the nonlinear system into an antagonistic player, yielding a conservative \textit{envelope} 
of the true HJR value function \cite{sharpless2024conservative}. This approach provides the necessary guarantees for safety-critical systems, but tends to be overly conservative for long time horizons or highly nonlinear systems. 

In this work, we generalize the results in \cite{sharpless2024conservative} to state-augmented (SA) systems. SA systems are popular for their ability to significantly outperform standard linearizations \cite{igarashi2020mpc}, and appear in, e.g., extended dynamic mode decomposition (EDMD) \cite{williams2015data},  learning-based linearizations \cite{yeung2019learning, CompKoop} and in other approximations of the Koopman Operator \cite{brunton2016koopman, abraham2019active, haseli2024modeling}. It is well-known that in the (asymptotic) limit of increasing dimension, the map of certain ``lifted'' models approaches the action of the Koopman Operator \cite{korda2018convergence, nuske2023finite, bramburger2023auxiliary} which exactly represents the nonlinear dynamics \cite{koopman1931hamiltonian}.

In this paper, we show that for any SA linearization one can define a game with antagonistic error such that its solution is guaranteed to be conservative with respect to the true solution of the nonlinear system (Theorem ~\ref{thm:HJR-SAS}). The proof follows from a series of value function comparisons and may be found in Sec.~\ref{sec:appen}. An immediate corollary is that reachability in the SA linear game implies reachability in the true system and, hence, the optimal control policy derived from the corresponding SA HJ value function is guaranteed to succeed in the original nonlinear system despite any disturbance or error of the SA model \eqref{HJR-SAS-ctrl}. In Sec.\ref{sec:SM}, we demonstrate the result in detail with the slow-manifold system, well known for having an exact SA representation in the autonomous case; in the inexact case with inputs, we show the augmented value function offers conservative envelopes (Figure~\ref{fig:SM}). Finally, we demonstrate in the controlled Van der Pol system in Sec.\ref{sec:VdP} how these results may be applied with various lifting functions to observe their corresponding conservative envelopes of the true solution where the controller is guaranteed to succeed (Figure~\ref{fig:VDP}).




The contributions of this work include:

\begin{enumerate}

    \item a novel Hopf-amenable method for generating conservative envelopes of HJ value functions via a SA linear game with antagonistic error,

    \item a formal proof that the resulting controller and reachable set are conservative for both the performance (\textit{Reach}) and safety (\textit{Avoid}) problem formulations, and
    
    \item demonstrations with comparison to traditional DP-based HJR, and with various lifting functions.
    
\end{enumerate}


\section{Preliminaries}\label{sec:prelims}

This paper focuses on control-affine and disturbance-affine systems of the form
\begin{eqnarray}
\dot x = f_x(x) + h_1(x)u + h_2(x)d \triangleq f(x, u, d)
\label{Dynamics}
\end{eqnarray}
with state $x \in \ssp \triangleq \mathbb{R}^\dimst$, and control $u$ and disturbance $d$ drawn from compact \& convex sets $\cset \subseteq \mathbb{R}^{n_u}$, $\dset \subseteq \mathbb{R}^{n_d}$. Let the input signals $u(\cdot)$ and $d(\cdot)$ be drawn from $\csetsig  \triangleq \{ \nu:\tint \mapsto \cset \mid \nu \text{ measurable} \}$ and $\dsetsig \triangleq \{ \nu:\tint \to \dset \mid \nu \text{ measurable} \}$.
Let $f$ be Lipschitz continuous in $(x,u,d)$ s.t. there exists a unique trajectory $\tj: \tint \to \ssp$ of $f$ defined by $\tj(\ti) = x$ and $\dot{\tj}(\tau) = f(\tj(\tau), u(\tau), d(\tau))$ a.e. for $\tau \in \tint$. For clarity, we at times write $\tj(\tau; x, u(\cdot), d(\cdot), t)$.


\subsection{Hamilton-Jacobi Reachability Problem}\label{sec:HJR}

To design a safe autonomous controller, HJR solves the optimal control that counters an adversarial disturbance in a differential game. 
The game is defined by the cost functional,
\begin{eqnarray}
\cost(x, u(\cdot), d(\cdot), t) \triangleq \tgfn(\tj(T)) + \int_t^T L(u(\tau), d(\tau)) d\tau ,
\label{GameCost}
\end{eqnarray}
which scores a trajectory for given input signals. Let the \textit{Reach} game be defined as the problem where the objective of Player I, the control, is to minimize (\ref{GameCost}) while the objective of Player II, the disturbance, seeks to maximize it. Let the \textit{Avoid} game be defined s.t. the player objectives are swapped. 

The terminal cost $\tgfn:\ssp \to \mathbb{R}$ is a convex, proper, lower semicontinuous function chosen such that
\begin{eqnarray}
\begin{cases}
\tgfn(x) < 0 \:\: \text{ for } \:\:x \in \tg \setminus \partial \tg \\
\tgfn(x) = 0 \:\: \text{ for }\:\:x \in \partial \tg \\
\tgfn(x) > 0 \:\: \text{ for } \:\:x \notin \tg \\
\end{cases}
\label{TargetValue}
\end{eqnarray}
where $\tg \in \ssp$ is a user-defined, closed set representing the target to reach or avoid and $\partial \tg$ is its boundary. To allow any feasible inputs to guide or perturb the trajectory, we assume the running cost $L \equiv 0$ in this work, but the derived conservative guarantee of the value may apply for any convex $L$ \cite{darbon2016algorithms, chow2017algorithm} but no longer characterizes reachability.  
In this context, we have defined the cost such that it has the property, 
\begin{eqnarray}
\begin{aligned}
\cost(x, u(\cdot), d(\cdot), t) \leq 0 &\iff \tgfn(\tj(T) \le 0 \\ &\iff \tj(T) \in \tg,  
\end{aligned}
\label{GameMeaning}
\end{eqnarray}
where $\tj(T) = \tj(T; x, u(\cdot), d(\cdot), t)$. 

Consider the game in which the disturbance has an instantaneous information advantage, but plays with respect to previous observations only. Formally, let a strategy $\strd: \csetsig \to \dsetsig$ for Player II be drawn from the set of non-anticipative strategies \cite{bacsar1998dynamic, evans1984differential, mitchell2005time}, 
\begin{eqnarray}
\begin{aligned}
\strd \in \stratset \triangleq \{ \gamma \mid u(\tau) = \hat u (\tau), \text{ a.e. } \tau \in \tint \implies \\ \gamma[u](\tau) = \gamma[\hat u] (\tau), \text{ a.e. } \tau \in \tint \}.
\label{NonAntStrats}
\end{aligned}
\end{eqnarray}
Then the value functions  $V, V^-:\ssp \times (-\infty,T] \to \mathbb{R}$ corresponding to the values of the \textit{Reach} and \textit{Avoid} games resp. are defined as 
\begin{eqnarray}
\begin{aligned}
V(x,t) & \triangleq \sup_{\strd \in \stratset} \inf_{u(\cdot) \in \csetsig} \tgfn(\tj(T; x, u(\cdot), \strd[u](\cdot), t)),
\\
V^-(x,t) & \triangleq \inf_{\strd \in \stratset} \sup_{u(\cdot) \in \csetsig} \tgfn(\tj(T; x, u(\cdot), \strd[u](\cdot), t)).
\end{aligned}
\label{GameValue}
\end{eqnarray}
At times, we write $V(x,t; f, \tg)$ to clarify this definition. Analogous to (\ref{GameMeaning}), these functions have the sublevel property,
\begin{eqnarray}
\begin{aligned}
V(x,t) &\leq 0 \iff x \in \hjrset(\tg,t), \\
V^-(x,t) &\leq 0 \iff x \in \hjrset^-(\tg,t)
\label{ValueMeaning}
\end{aligned}
\end{eqnarray}
where $\hjrset(\tg,t)$ \& $ \hjrset^-(\tg,t)$ are the backward reachable sets (BRSs): the set of states which may be driven to the target at time $T$ (starting from time $t$) despite any disturbance (\textit{Reach }set) or despite any control (\textit{Avoid }set) respectively,
\begin{eqnarray}
\begin{aligned}
\hjrset(\tg,t) \triangleq \{x \mid \forall 
\strd \in \stratset, \exists u(\cdot) \in \csetsig  \text{ s.t. } 
\\ 
\tj(T; x, u(\cdot), \strd[u](\cdot), t) \in \tg \},
\\
\hjrset^-(\tg,t) \triangleq \{x \mid \exists \strd \in \stratset, \forall u(\cdot)\in \csetsig \text{ s.t. } 
\\ 
\tj(T; x, u(\cdot), \strd[u](\cdot), t) \in \tg \}.
\end{aligned}
\label{BRS}
\end{eqnarray}
In contrast, consider the set of all backwards \textit{feasible} states $\fset \subseteq \ssp$ for which there exist bounded input signals that \textit{could} drive the trajectory into the target at time $T$ starting from time $t$, given by
\begin{eqnarray}
\begin{aligned}
\fset (\tg,t) \triangleq \{x \mid \exists u(\cdot) \in \csetsig, \quad \exists d(\cdot) \in \dsetsig \text{ s.t. } 
\\ 
\tj(T; x, u(\cdot), d(\cdot), t) \in \tg \}.
\end{aligned}
\end{eqnarray}
By Filippov and others, this set will be bounded for any compact sets $\tg$, $\cset$, $\dset$ and Lipschitz dynamics \cite{filippov1960differential, aubin2012differential, scott2013bounds}. Additionally, the backwards feasible tube $\ftube$ will be a relevant object for bounding trajectories of the game and is given by
\begin{eqnarray}
\begin{aligned}
\ftube(\tg, t) \triangleq \bigcup_{\tau \in \tint} \fset(\tg,\tau),
\label{rsetdef}
\end{aligned}
\end{eqnarray}
which we may also know is bounded given the above assumptions for the closed interval $\tint$. In antagonistic or worst-case scenarios, $\fset$ \& $\ftube$ are insufficient for guaranteeing $\tj(T) \in \tg$, however, they may be used for bounding trajectories in order to define an antagonistic error player \cite{sharpless2024conservative}, which offers a conservative guarantee for a linear game with respect to a nonlinear game.

Notably, applying Bellman's principle of optimality to the value function $V$ leads to the following well-known theorem. 

\begin{Theorem}
[Evans 84] \cite{evans1984differential} \\
Given the assumptions (2.1)-(2.5) in [Evan 84], the value function $V$ defined in (\ref{GameValue}) is the viscosity solution to the following Hamilton-Jacobi Partial Differential Equation,
\begin{eqnarray}
\begin{aligned}
\dot V  + \ham(x, \nabla_x V, \tau) & = 0 &\text{ on } \ssp& \times \tint, \\
V(x,T) & = \tgfn(\tj(T)) &\text{ on } \ssp& 
\end{aligned}
\label{HJPDE-V}
\end{eqnarray}
where the Hamiltonian $\ham:\ssp \times \ssp \times \tint \to \mathbb{R}$ is 
\begin{eqnarray}
\ham(x, p, t) = \minou \maxod p \cdot f(x, u, d).
\label{Hamiltonian}
\end{eqnarray}
\label{thm:HJPDE}
\end{Theorem}

This equivalently applies to $V^-$, but note that the Hamiltonian in the \textit{Avoid} game takes the flipped form $H^-(x,p,t)=\max_u \min_d p \cdot f(x,u,d)$. In either game, solving this PDE yields the value function and corresponding BRS. Additionally, the value function can be used to derive the optimal control policy for $(x,t)$, e.g., for the \textit{Reach} game:
\begin{eqnarray}
\begin{aligned}
u^*(\tau) = \arg \minou \nabla_x V(\tj(\tau; x),\tau) \cdot h_1(\tau)u.
\end{aligned}
\label{HJoc}
\end{eqnarray}
The main challenge of HJR lies in solving the PDE in (\ref{HJPDE-V}); DP methods propagate $V(x, t)$ by finite-differences over a grid of points that grows exponentially with respect to $\dimst$ \cite{bansal2017hamilton}. In practice, this is computationally intractable for systems with dimension $\dimst \geq 6$ and is constrained to offline analysis.

Notably, it has been shown in \cite{kurzhanski2014dynamics, darbon2016algorithms, chow2017algorithm} that if a system has linear dynamics and the target is convex, then the generalized Hopf formula \cite{rublev2000generalized} gives the viscosity solution of (\ref{HJPDE-V}). Hence, instead of DP, the value may be feasibly solved by optimization of the Hopf formula independently over space, and this has been demonstrated for systems of up to dimension $\dimst = 4096$ \cite{chow2018algorithm}. However, this is limited to linear dynamics and motivates the current work.

\subsection{State Augmented Systems}\label{sec:SA}

Consider an augmentation of the space $\ssp$, say $\sasp \triangleq \mathbb{R}^{\dimsa}$. Namely, let the \textit{lifting function} $\lf:\ssp \to \sasp$ be a bounded map from the state space to the augmented space that takes the following form,
\begin{eqnarray}
\lf(x) \triangleq [x, \lfl_1(x), \dots, \lfl_{\dimsa - \dimst}(x)]^\top, \lfl_i: \ssp \to \mathbb{R}
\label{LiftingFn}
\end{eqnarray}
where $\lfl_i \in C^1$ are smooth, user-defined functions that are chosen to improve the linearization accuracy, e.g. a truncated functional basis. The range $\rglf \subseteq \sasp$ represents a manifold in the augmented space (see, for example, Figure~\ref{fig:SM}). By definition, $\lf$ is injective and therefore has an inverse in the range, say $\lfi: \rglf \to \ssp$ with $\lfi (g) \triangleq x$ if $g = \lf(x)$. 

Let the map $P: \sasp \to \ssp$ be the projection of the augmented space onto the state space, which in this context takes the form of a matrix $P = [I_\dimst \: 0_{\dimsa-\dimst}]$. By definition, 
\begin{eqnarray}
x = \pj \lf(x), \text{ hence } \lfi = \pj|_{\rglf}
\label{Proj}
\end{eqnarray}
where $\pj|_{\rglf}: {\rglf} \to \ssp$ is the restriction of the map to the manifold.

Additionally, consider a linear model in $\sasp$, 
\begin{eqnarray}
    \dot g \approx \fglin (g, u, d) \triangleq Kg + L_1 u + L_2 d.
    \label{fglin}
\end{eqnarray}
where $K, L_1, L_2 \in \mathbb{R}^{\dimsa \times \dimsa}, \mathbb{R}^{\dimsa \times n_u}, \mathbb{R}^{\dimsa \times n_d}$ are real matrices. This system may be generated in a variety of ways, including, e.g., via the taylor series or least-squares fitting. The principal result will hold for any finite linear model, since a finite linear map is bounded, yielding a finite maximum error on the bounded set $\ftube$ \cite{sharpless2024conservative}. In the original space, this error may be large, giving overly conservative envelopes for long time-horizons or high nonlinearity, but in SA systems, it is well-known that in the limit of increasing dimension, there are linear models whose output will approach the action of the Koopman operator asymptotically, and thus, the error tends to zero \cite{korda2018convergence, nuske2023finite, bramburger2023auxiliary}.

\section{RESULTS}\label{sec:SfEnv}

\begin{figure*}[t] 
    \centering
    \includegraphics[width=\linewidth]{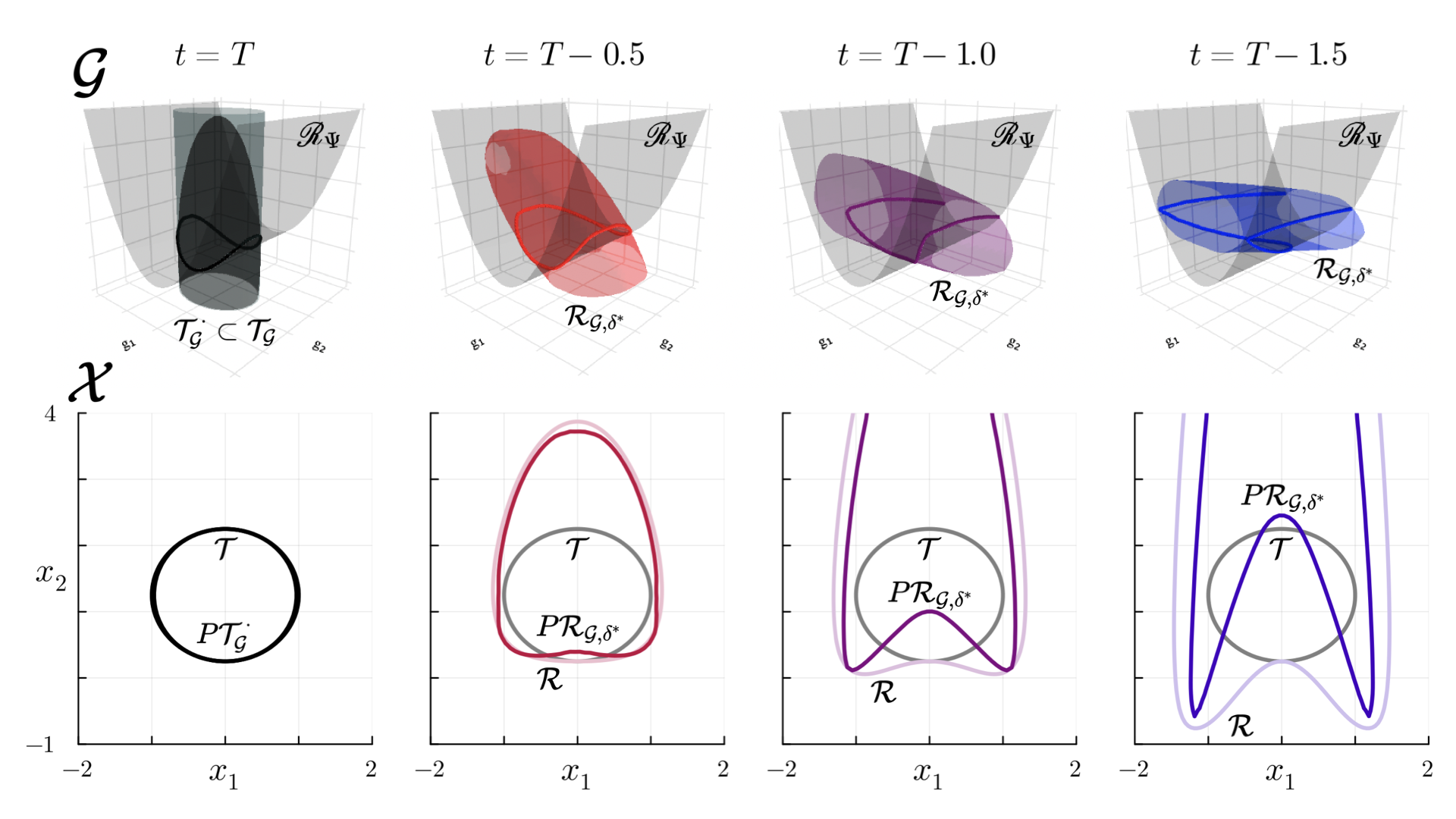}    
    \caption{\textbf{\textit{Reach} game in the Slow Manifold system (Sec.\ref{sec:SM}).}
   The bottom left shows the target $\tg \subset \ssp$. The corresponding augmented target $\tgg \subset \sasp$, inner augmented target $\tgg^\inn \subset \tgg$ and manifold $\rglf$ are shown in the top left. The projection of $\tgg^\inn$ restricted to the manifold $\pj \tgg^\inn \subset \tg$ is also shown in the bottom left. The reach set $\hjrset_{\sasp, \maxerr}(\tgg^\inn,t)$ of the SA linear game with antagonistic error is shown in the top row for different time horizons $t$ (colored transparent sets), and its projection restricted to the manifold $\pj \hjrset_{\sasp, \maxerr}(\tgg^\inn,t)$ is shown in the bottom row (dark-colored lines). By Theorem \ref{thm:HJR-SAS}, $\pj \hjrset_{\sasp, \maxerr}(\tgg^\inn,t) \subset \hjrset$, the reach set $\hjrset$ of the nonlinear game (light-colored lines).
   }
    \label{fig:SM}\vspace{-1em}
\end{figure*}

In this section, we state the main theoretical result, namely that the true nonlinear game value may be conservatively approximated by a linear game value with antagonistic error in the state augmented system. Toward defining this latter game, let the \textit{augmented target} $\tgg \subseteq \sasp$ be defined as,
\begin{eqnarray}
    \tgg \triangleq \{ g \in \sasp \mid \pj g \in \tg \}.
    \label{AugTarget}
\end{eqnarray}
Informally, this definition extrudes the target over the augmented variables in an indiscriminate manner (see the upper left panel of Figure~\ref{fig:SM}). By the assumption that $\tg$ is closed, it follows that $\tgg$ is also closed. For general nonlinear $f$, the conservative guarantee we will show also requires boundedness of the target, hence consider any compact sets $\tgg^\inn, \tgg^\out \subset \sasp$ satisfying $(\tgg^\inn \cap \rglf ) \subseteq \rglftg$ and $(\tgg^\out \cap \rglf ) \supseteq \rglftg$. Informally, these sets suffice as inner and outer bounds of $\tgg$ for trajectories invariant to the manifold (Lemma ~\ref{lem:valgbound}). Let their terminal costs $\tgfg^\inn$ and $\tgfg^\out$ be defined as in (\ref{TargetValue}).

To bridge the games, we also make use of the following nonlinear dynamics in $\sasp$,
\begin{eqnarray}
    \dot g = \nabla_x \lf(\pj g) \cdot f(\pj g,u,d)  \triangleq  \fg (g, u, d).
    \label{fg}
\end{eqnarray}
By the assumptions on $f$ and $\lf$, this system will be Lipschitz and bounded. 
Hence, for any bounded $\fseto$ s.t. $\fseto \subset \sasp$ the maximum error between $\fg$ and $\fglin$ given by
\begin{eqnarray}
    \delta^*(\fseto) \triangleq \sup_{\fseto \times \cset \times \dset} \bigg\Vert [\fg - \kappa](g,u,d) \bigg\Vert
\end{eqnarray}
is finite. The novelty in the present work is recognizing that with $\fg$ it is possible to generalize previous conservative linearization results \cite{sharpless2024conservative} to the augmented space where the error $\maxerr$ may be smaller with a high-dimensional lift \cite{korda2018convergence}.

We may now consider the principal result.

\begin{Theorem} \label{thm:HJR-SAS}
    Let $\Vge$ \& $\Vge^-$ be the viscosity solutions of
    \begin{eqnarray}
    \begin{aligned}
    &\dot{V}_{\sasp, \maxerr} + \hamlerr(g, \nabla_g \Vge, t) = 0, \Vge(g,T) = \tgfg^\inn(g),\\
    &\dot{V}_{\sasp, \maxerr}^- + \hamlerr^-(g, \nabla_g \Vge^-, t) = 0, \Vge^- (g,T) = \tgfg^\out(g),
    \end{aligned}
    \label{HJPDE-Vge}
    \end{eqnarray}
    where $\hamlerr$ and $\hamlerr^-$ are defined by
    \begin{eqnarray}
    \begin{aligned}
    \hamlerr(g, p, t) \triangleq \minou \maxod \max_{\eps \in \errset_\sasp} p \cdot (\kappa(g,u,d) + \eps), \\
    \hamlerr^-(g, p, t) \triangleq \maxou \minod \min_{\eps \in \errset_\sasp} p \cdot (\kappa(g,u,d) + \eps).
    \end{aligned}
    \label{AugLinearHamiltonianEps}
    \end{eqnarray}
    with $\errset_\sasp \triangleq \errset(\maxerr_\sasp)$. Then in the \textit{Reach} and \textit{Avoid} games, if 
    $\maxerr_\sasp \triangleq \maxerr(\rglffeas)$, it follows $\forall x \in \ftube(\tg, t)$,
    \begin{eqnarray}
    \begin{aligned}
    \Vge(\lf(x), t) \le 0 &\implies V(x, t) \le 0 , \\
    \Vge^-(\lf(x), t) > 0 &\implies V^- (x, t) > 0.
    \end{aligned}
    \end{eqnarray}
    Moreover, if $\lf(x) \in \hjrset_{\sasp, \maxerr}(\tgg^\inn, t)$ in the \textit{Reach} game or $\lf(x) \notin \hjrset^-_{\sasp, \maxerr}(\tgg^\out, t)$ in the \textit{Avoid} game, the optimal policies $u^*_{\sasp, \maxerr}(\cdot)$ \& $u^{*,-}_{\sasp, \maxerr}(\cdot)$ resp. will for any $\strd \in \stratset$ yield,
    \begin{eqnarray}
    \begin{aligned}
    \tj(T; x, u^*_{\sasp, \maxerr}(\cdot), \strd[u^*_{\sasp, \maxerr}](\cdot), t) \in \tg, \\
    \tj(T; x, u^{*,-}_{\sasp, \maxerr}(\cdot), \strd[u^{*,-}_{\sasp, \maxerr}](\cdot), t) \notin \tg.
    \end{aligned} \label{HJR-SAS-ctrl}
    \end{eqnarray}
\end{Theorem}
The proof is in Sec.~\ref{sec:appen}.  
Intuitively, Theorem~\ref{thm:HJR-SAS} seeks to show that the SA linear game with error will be conservative w.r.t. the nonlinear game, yielding a controller that is guaranteed to win in $\ssp$ when it can in $\rglf$. Notably, since the error of the SA linear dynamics vanishes with increasing dimension, the SA envelope will thus \textit{conservatively} approach the true value.
The proof is challenging because in the SA space, the linear trajectories are not invariant to the manifold $\rglf$, a well-known issue in EDMD  \cite{bruder2019modeling}. Previous work has attempted to project the invariant evolution, either via $\ssp$ first or directly to $\rglf$, both of which involve nonlinear operations and hence corrupt the purely linear evolution.

To overcome this challenge, the proof of Theorem ~\ref{thm:HJR-SAS} relies on a sequence of value comparisons. In Lemmas~\ref{lem:ptj} \& \ref{lem:valffg}, we show that the relationship between trajectories of $f$ and $\fg$, $$\tj(\tau; x, u(\cdot), d(\cdot), t) = P \tjg(\tau; \lf(x), u(\cdot), d(\cdot), t),$$ yields an equivalence between value of the original game at a state $x$ and that of a game in the SA space with $\fg$ and $\tgg$ at the augmentation of the state $\lf(x)$, $$V(x,t; f, \tg) = \Vg(\lf(x),t; \fg, \tgg).$$ Second, given that the bounded sets  $\tgg^\inn$ and $\tgg^\out$ are covered by and cover all feasible endpoints $\tjg(T)$ in $\tgg$ resp., we next prove (Lemma~\ref{lem:valgbound})
\begin{eqnarray*}
\begin{aligned}
\Vg(g, t; \fg, \tgg^\inn) \le 0 &\implies \Vg(g, t; \fg, \tgg)  \le 0, \\
\Vg^-(g, t; \fg, \tgg^\out) > 0 &\implies \Vg^-(g, t; \fg, \tgg) > 0 .
\end{aligned}
\end{eqnarray*}
Finally, it is then possible to apply Theorem 3 of \cite{sharpless2024conservative} to generate an envelope of the nonlinear SA game with a bounded target by transforming the error between $\fg$ and $\kappa$ on the backwards feasible tube mapped to the SA space into an antagonistic player (Corollary \ref{lem:coro}). For the guarantee, the antagonistic error needs only to be capable of inducing the trajectories of $\fg$, which are invariant to $\rglf$, hence, \textit{error off the manifold is irrelevant}. Let 
\begin{equation}
    \pj \hjrset_\sasp \triangleq \{ x \in \ssp \mid x = Pg, g \in \hjrset_\sasp \cap \rglf \}
\end{equation}
represent the projection (restricted to the manifold) of any augmented reachable set. Then the above sequence may be understood equivalently in \textit{Reach} and \textit{Avoid} games as,
\begin{align*}
    &\pj \hjrset_{\sasp, \maxerr}(\tgg^\inn, t) \subseteq \pj \hjrset_{\sasp}(\tgg^\inn, t) \subseteq \pj \hjrset_{\sasp}(\tgg, t) = \hjrset(\tg, t), \\
    &\pj \hjrset_{\sasp, \maxerr}^-(\tgg^\out, t) \supseteq \pj \hjrset^-_{\sasp}(\tgg^\out, t) \supseteq \pj \hjrset^-_{\sasp}(\tgg, t) = \hjrset^-(\tg, t).
\end{align*}

In summary, solving the value for a linear game with antagonistic error in the state-augmented space on the manifold offers a safe approximation of the true value and yields an optimal controller that rejects any disturbance in the true dynamics or error from the approximation. Conservative, convex approximations of $\ftube$ may be solved rapidly with Differential Inclusion methods \cite{aubin2012differential, chen2012taylor} upon which $\maxerr$ may be computed. Moreover, there are several additional corollaries which may be extended from \cite{sharpless2024conservative} for reducing $\maxerr$ such as via ensembles, partitions, and \textit{forward} feasible sets, but we leave this to future work. Ultimately, Theorem~\ref{thm:HJR-SAS} is meaningful because the error $\maxerr$ may be smaller for a high-dimensional SA model \cite{korda2018convergence, nuske2023finite, bramburger2023auxiliary}, yielding a safe linear game of improved accuracy, i.e. reduced conservativeness, that may yet be solved by the Hopf formula with vastly improved speed and dimensionality limitation.

Interestingly, unlike previous linear methods, this result allows the safe approximation of HJR sets in a non-convex fashion. It is well-known that for a linear game with a convex target, the level sets must remain convex \cite{kurzhanski2014dynamics} (see the colored sets in the top row of Figure~\ref{fig:SM}). However, by comparing games across the nonlinear map $\lf$, this restriction may be circumvented: the game value level sets on the range of $\lf$, i.e. the intersection of the convex solutions with the nonlinear manifold, may be non-convex (see $\pj \hjrset_{\sasp, \maxerr}$ in Figure~\ref{fig:SM}). 

\section{Proof of Theorem \ref{thm:HJR-SAS}} \label{sec:appen}

We now prove Theorem ~\ref{thm:HJR-SAS} and the lemmas necessary for it. We begin by proving a valuable relation between trajectories of $f$ and $\fg$.

\begin{Lemma} (Equivalence of Projected Trajectories for $\fg$) \label{lem:ptj} \\
Let $\tjg: \tint \to \sasp$ be a trajectory of (\ref{fg}) s.t. $\tjg (t) = g$ for $g \in \sasp$ 
and $\dot{\tj}_\sasp (\tau) = \fg(\tjg (\tau), u(\tau), d(\tau))$. Then $\forall \tau \in \tint$, $u(\cdot) \in \csetsig$, $d(\cdot) \in \dsetsig$, if $x = \pj g$,
\begin{eqnarray}
\pj \tjg (\tau; g, u(\cdot), d(\cdot), t) = \tj(\tau; x, u(\cdot), d(\cdot), t).
\end{eqnarray}
\end{Lemma}
\begin{proof}
The proof is an extension of the standard ODE uniqueness proof under Lipschitz condition \cite{boyce2021elementary}. Recall, a trajectory is given implicitly by $\tj(\tau) = x + \int_t^\tau f(\tj(s), u(s), d(s))\,ds$. 
Then, since $\pj$ is linear, 
\begin{equation*}
\begin{aligned}
    \pj \tjg (\tau) & = \pj g + \int_t^\tau \pj \fg (\tjg(s), u(s), d(s)) \, ds \\
& = x + \int_t^\tau f ( \pj \tjg(s), u(s), d(s)) \, ds  
\end{aligned}
\end{equation*}
where the second line follows from the definition of $\lf$. Then, at time $\tau$,
\begin{eqnarray*}
    \begin{aligned}
        \Vert &\pj \tjg(\tau) - \tj(\tau) \Vert \\ 
        &= \bigg\Vert \int_t^\tau f (\pj \tjg (s), u(s), d(s)) - f(\tj(s), u(s), d(s)) \, ds \bigg \Vert \\
        &\le L_f \int_t^\tau \Vert \pj \tjg (s) - \tj(s) \Vert \, ds
    \end{aligned}
\end{eqnarray*}
where $L_f$ is the Lipschitz constant for $f$. Writing $\phi(\tau) = \int_t^\tau \Vert \pj \tjg (s) - \tj(s) \Vert \, ds$, then we directly have $\dot \phi - L_f \phi \le 0, \phi \ge 0, \phi(t) = 0$, and the Gronwall inequality gives $ 0 \le \phi(\tau) \le \phi(t) \exp( L_f t ) = 0 $ and therefore $\phi(\tau) \equiv 0$.
\end{proof}

With this result, we may show the equivalence of the games defined with $f$ \& $\tg$ and $\fg$ \& $\tgg$.

\begin{Lemma} (Equivalence of Value for $\fg$) \label{lem:valffg} \\ 
Let $\tgg \triangleq \{ g \mid \pj g \in \tg \}$ with $\tgfg(g) \triangleq \tgfn (\pj g)$. Then if the \textit{Reach} and \textit{Avoid} game values are defined analogous to (\ref{GameValue}) for $\tgg$ and $\fg$ s.t.
\begin{eqnarray}
\begin{aligned}
\Vg(g,t) &\triangleq \sup_{\strd \in \stratset} \inf_{u(\cdot) \in \csetsig} \tgfg (\tjg (T; g, u(\cdot), \strd[u](\cdot), t)) \\
\Vg^-(g,t) &\triangleq \inf_{\strd \in \stratset} \sup_{u(\cdot) \in \csetsig} \tgfg (\tjg (T; g, u(\cdot), \strd[u](\cdot), t)),
\end{aligned}
\end{eqnarray}
then it must hold that for any $x = \pj g$,
\begin{eqnarray}
\begin{aligned}
\Vg(g, t; \fg, \tgg)  &= V(x, t; f, \tg), \\
\Vg^-(g, t; \fg, \tgg) &= V^-(x, t; f, \tg),
\end{aligned}
\end{eqnarray}
and moreover, the optimal strategies are equivalent.
\end{Lemma}
\begin{proof}
We will prove the result for the \textit{Reach} game which is identical to the \textit{Avoid}. Consider the trajectory $\tjg(\tau)$ with initial state $g$ arising from $u (\cdot) \in \csetsig$ and $d(\cdot) \in \dsetsig$. By definition the cost of this trajectory will be,
$$
\tgfg ( \tjg(\tau)) = \tgfn ( \pj \tjg(\tau)),
$$
and by Lemma \ref{lem:ptj}, $\forall \tau \in \tint, \pj \tjg (\tau) = \tj(\tau)$, thus
$$
\tgfg ( \tjg(\tau)) = \tgfn (\tj(\tau)), \quad \tj(t) = x = Pg.
$$
It follows that for $\tau = T$, $x = Pg$,
\begin{eqnarray}
\begin{aligned}
    \sup_{\strd \in \stratset}& \inf_{u(\cdot) \in \csetsig} \tgfg (\tjg (T; g, u(\cdot), \strd[u](\cdot), t)) \\& = \sup_{\strd \in \stratset} \inf_{u(\cdot) \in \csetsig} \tgfn(\tj(T; x, u(\cdot), \strd[u](\cdot), t)) \\ \implies & \Vg(g, t; \fg, \tgg) = V(x, t; f, \tg), 
\end{aligned}
\end{eqnarray}
and because the objectives and argument spaces are identical, it must be that the optimizing arguments are equivalent.
\end{proof}

We would like to now use this nonlinear game with $\fg$ to generate a safe envelope with the linear system and bounded error as in \cite{sharpless2024conservative}. However, in order to apply this to nonlinear dynamics that are not bounded absolutely, it is necessary to consider bounded sets of the trajectories i.e. the feasible tube, and hence a bounded target is required.

\begin{Lemma} (Conservative, Bounded Augmented Sets) \label{lem:valgbound} \\
Let $\tgg^\inn, \tgg^\out \subset \sasp$ be any closed, bounded sets satisfying $(\tgg^\inn \cap \rglf ) \subseteq \rglftg$ and $(\tgg^\out \cap \rglf ) \supseteq \rglftg$, which define $\tgfg^\inn$ \& $\tgfg^\out$ as in (\ref{TargetValue}). Then in the \textit{Reach} and \textit{Avoid} games, $\forall g \in \rglf$,
\begin{eqnarray}
\begin{aligned}
\Vg(g, t; \fg, \tgg^\inn) \le 0 &\implies \Vg(g, t; \fg, \tgg)  \le 0, \\
\Vg^-(g, t; \fg, \tgg^\out) > 0 &\implies \Vg^-(g, t; \fg, \tgg) > 0 .
\end{aligned}
\end{eqnarray}
\end{Lemma}
\begin{proof}
Note, by definition of the augmented target, $\rglftg = \tgg \cap \rglf$. Hence, the assumptions on $\tgg^\inn$ and $\tgg^\out$ imply that, $\forall g \in \rglf$, then $\tgfg^\inn(g) \le  0 \implies \tgfg(g) \le 0$ and $\tgfg^\out(g) > 0 \implies \tgfg(g) > 0$. The \textit{Reach} and \textit{Avoid} proofs are mirrored hence we will show only the \textit{Avoid} for brevity. 

For contradiction, assume $\exists g \in \rglf$ s.t. $\Vg^-(g, t; \fg, \tgg^\out) > 0$ but $\Vg^-(g, t; \fg, \tgg) \le 0$. If
$$ \Vg^-(g, t; \fg, \tgg^\out) = \inf_{\strd} \sup_{u(\cdot)} \tgfg^\out ( \tjg(T; g, u(\cdot), \strd[u](\cdot),t)) > 0,$$
then $\exists \epsilon > 0, \forall \strd \in \stratset$ s.t.
$$\sup_{u(\cdot)} \tgfg^\out ( \tjg(T; g, u(\cdot), \strd[u](\cdot),t)) > 2 \epsilon > 0$$
and thus, $\exists \epsilon > 0, \forall \strd \in \stratset, \exists u(\cdot) \in \csetsig $ s.t.
$$\tgfg^\out ( \tjg(T; g, u(\cdot), \strd[u](\cdot),t)) > \epsilon > 0.$$
But, $\forall g \in \rglf, \tgfg^\out (g) > 0 \implies \tgfg (g) > 0$, hence, $\exists \epsilon, \forall \strd \in \stratset, \exists u(\cdot) \in \csetsig $ s.t.
$$\tgfg ( \tjg(T; g, u(\cdot), \strd[u](\cdot), t)) > \epsilon > 0. $$
Then,
$$\Vg^-(g, t; \fg, \tgg) = \inf_{\strd} \sup_{u(\cdot)} \tgfg ( \tjg(T; g, u(\cdot), \strd[u](\cdot),t)) > 0$$
which is a contradiction.
\end{proof}

It is now possible to apply the results of \cite{sharpless2024conservative} to create an envelope of the value with $\fg$ with the linear model $\fglin$ with antagonistic error $\varepsilon$.

\begin{Corollary} (Conservative Linearization) \label{lem:coro} \\
Let the maximum error $\maxerr_\sasp$ define the set of measurable functions $\errsigset: \tint \to \errset(\maxerr_\sasp)$, and non-anticipative strategies $\strateset: \tint \to \errsigset$. Then if the \textit{Reach} and \textit{Avoid} game values are defined analogous to (\ref{GameValue}) s.t.
\begin{eqnarray}
\begin{aligned}
&\Vge(g,t) \triangleq \\ &\sup_{\stre \in \strateset}\sup_{\strd \in \stratset} \inf_{u(\cdot) \in \csetsig} \tgfg^\inn (\tjgk (T; g, u(\cdot), \strd[u](\cdot), \stre[u](\cdot), t)), \hspace{-20pt} \\
&\Vge^-(g,t) \triangleq \\ &\inf_{\stre \in \strateset}\inf_{\strd \in \stratset} \sup_{u(\cdot) \in \csetsig} \tgfg^\out (\tjgk (T; g, u(\cdot), \strd[u](\cdot),\stre[u](\cdot), t)), \hspace{-20pt}
\end{aligned}
\label{Vgedefs}
\end{eqnarray}
where $\tjgk$ are trajectories of the dynamics $\fglin(g,u,d) + \eps$. Then in the \textit{Reach} and \textit{Avoid} games, if $\maxerr_\sasp \triangleq \maxerr(\rglffeas)$, it follows $\forall g \in \rglffeas$,
\begin{eqnarray}
    \begin{aligned}
    \Vge(g, t; \fglin + \varepsilon, \tgg^\inn) \le 0 & \implies \Vg(g, t; \fg, \tgg^\inn) \le 0 , \\
    \Vge^-(g, t; \fglin + \varepsilon, \tgg^\out) > 0 &\implies \Vg^- (g, t; \fg, \tgg^\out) > 0.
    \end{aligned}
\end{eqnarray}
Moreover, the optimal strategies reject the true error.
\end{Corollary}
\begin{proof}
    To apply Theorem 3 in \cite{sharpless2024conservative}, we must show
    \begin{eqnarray*}
    \begin{aligned}
    \rglffeas \supseteq \{y \in \sasp \mid y& = \tjg(\tau; g, u(\cdot), d(\cdot), t), \tjg(T) \in \tgg, \\ &u(\cdot) \in \csetsig, d(\cdot) \in \dsetsig \}.
    \end{aligned}
    \end{eqnarray*}
    By Lemma \ref{lem:ptj}, for any $y$ in the RHS set, $\pj y=\pj\tjg(\tau) = \tj(\tau)$ and at $\tau = T$, by definition of $\tgg$ this implies $\pj\tjg(T) \in \tg$. Hence, $\pj y \in \ftube(\tg, t)$. Since $g=\Psi(x)$, by definition $\tjg(\tau) \in \rglf$. This implies $\lf(\pj y) = y$. Well, $\rglffeas 
    \triangleq \{\lf(x), x \in \ftube(\tg, t)\}$ hence $y \in \rglffeas$.
\end{proof}
    
Informally, this follows because the antagonistic error player draws from a set containing the true error and thus may always induce the true trajectory when it benefits them. Then the $\sup$ or $\inf$ over error strategies bounds the true game value \cite{sharpless2024conservative}. Finally, we prove Theorem \ref{thm:HJR-SAS}.

\begin{proof}

    First, with the compactness of $\tgg^\inn, \tgg^\out, \cset, \dset$ and the Lipschitz nature of $\kappa + \varepsilon$, the assumptions (2.1)-(2.4) of \cite{evans1984differential} are satisfied and $\Vge$ and $\Vge^-$ defined in (\ref{Vgedefs}) are the viscosity solutions of the HJ-PDEs given in (\ref{HJPDE-Vge}) by Theorem ~\ref{thm:HJPDE} \cite{evans1984differential}. The proof of the \textit{Reach} game is akin to that of the \textit{Avoid} game, hence we present only the former case. 
    
    Let $g \in \rglffeas$ then $g = \lf(x), Pg = x$. Assume,
    $$\Vge(g, t; \fglin + \varepsilon, \tgg^\inn) \le 0.$$
    Then by Corollary ~\ref{lem:coro}, Lemma ~\ref{lem:valgbound}, and Lemma ~\ref{lem:valffg},
    \begin{eqnarray*}
    \begin{aligned}
        \implies \Vge(g, t; \fg, \tgg^\inn) &\le 0 \\
        \implies \Vg(g, t; \fg, \tgg) &\le 0 \\
        \implies V(x, t; f, \tg) &\le 0,
    \end{aligned}
    \end{eqnarray*}
    which proves the claim for this case.
    For $\pj \hjrset_\sasp \triangleq \{ Pg \mid g \in \hjrset_\sasp \cap \rglf \}$, this is equivalent to saying,
    $$\pj \hjrset_{\sasp, \maxerr}(\tgg^\inn, t) \subseteq \pj \hjrset_{\sasp}(\tgg^\inn, t) \subseteq \pj \hjrset_{\sasp}(\tgg, t) = \hjrset(\tg, t).$$
    Lastly, if $\forall \stre \in \strateset, \forall \strd \in \stratset, \exists u_{\sasp, \maxerr}^*(\cdot) \in \csetsig$ s.t. 
    $$\tgfg^\inn (\tjgk(T; g, u_{\sasp, \maxerr}^*(\cdot), \strd[u_{\sasp, \maxerr}^*](\cdot), \stre[u_{\sasp, \maxerr}^*](\cdot), t)) \le 0$$
    by the same logical sequence as above,
    \begin{eqnarray*}
    \begin{aligned}
        \implies \exists \stre, \stre[u_{\sasp, \maxerr}^*] = \varepsilon(\cdot) \text { s.t. } \\ \tgfg^\inn (\tjg (T; g, u_{\sasp, \maxerr}^*(\cdot), \strd[u_{\sasp, \maxerr}^*](\cdot), t) &\le 0 \\
        \implies \tgfg (\tjg(T; g, u_{\sasp, \maxerr}^*(\cdot), \strd[u_{\sasp, \maxerr}^*](\cdot), t) &\le 0 \\
        \implies \tgfn ( \pj \tjg(T; g, u_{\sasp, \maxerr}^*(\cdot), \strd[u_{\sasp, \maxerr}^*](\cdot), t) &\le 0 \\
        \implies \tgfn ( \tj (T; g, u_{\sasp, \maxerr}^*(\cdot), \strd[u_{\sasp, \maxerr}^*](\cdot), t) &\le 0.
    \end{aligned}
    \end{eqnarray*}
    Hence, the control signal $u_{\sasp, \maxerr}^*(\cdot)$, which may be solved from the linear program in (\ref{HJoc}) with $L_1$ \& $\Vge$ (given by the optimal argument of the Hopf formula), will drive the true trajectory into the target despite any disturbance.
\end{proof}

\section{DEMONSTRATION}\label{sec:Demos}

\subsection{Slow Manifold System} \label{sec:SM}

To illustrate Theorem \ref{thm:HJR-SAS}, consider the well-known ``slow-manifold'' system  \cite{brunton2016koopman, korda2018linear} with inputs,
\begin{align}
    \dot{x} =  
    \begin{bmatrix} \mu x_1 \\ \lambda (x_2 - x_1^2) \end{bmatrix} + u + d,
    \label{SlowManifoldControl}
\end{align}
with $\mu, \lambda := -0.05, -1$, control $u\in\cset$ and disturbance $d\in\dset$. In the autonomous case, this system has an exact linear representation in the state augmented space defined by  $g=\Psi(x) \triangleq [x_1, x_2, x_1^2]^\top$ with range $\rg_{\lf}$ given by the quadratic $g_3 = g_1^2$. With this lift, the exact SA dynamics $f_{\sasp}$ are given by,
\begin{align}
    \dot{g} = f_{\sasp} (g, u, d) = 
    \begin{bmatrix} \mu & 0 & 0 \\ 0 & \lambda & - \lambda \\ 0 & 0 & 2\mu \end{bmatrix} g + \begin{bmatrix} 1 & 0 \\ 0 & 1 \\ 2g_1 & 0 \end{bmatrix}(u + d).
    \label{SlowManifoldControlLifted}
\end{align}
Of course, when $u$ and $d$ are nontrivial, the presence of $g_1$ in the affine term makes $f_{\sasp}$ nonlinear. 

Consider a game governed by the dynamics (\ref{SlowManifoldControl}) in which the controller aims to drive the trajectory from an initial $(x,t)$ such that at time $T$ the trajectory is in $\tg \triangleq \{ y \in \ssp \mid (y - c_\tg)^\top (y - c_\tg) \le 1 \}$ centered at $c_\tg \triangleq [0, 1.25]$, while the disturbance aims to do the opposite (\textit{Reach} game). Let control and disturbance sets be given by $\cset \triangleq \{ \Vert u \Vert_2 \le 1/2 \}$ and $\dset \triangleq \{ \Vert d \Vert_2 \le 1/4 \}$. Choose $ \tgg^\inn \triangleq \{ (g - \lf(c_\tg))^\top Q (g - \lf(c_\tg)) \le 1 \} $ with $Q = \text{Diag}([1,1, \eta])$, where $\eta=1/15$.  Since $\eta > 0$, then $\tgg^\inn \subset \tgg$. Let the linear model $\fglin$ be defined as in (\ref{SlowManifoldControlLifted}) with $g_1 = \lf(c_\tg)_1 = 0$ in the input-affine term. The tube $\ftube(\tg, t)$ is conservatively solved (via  \cite{JuliaReach19}) and on a grid over $\ftube(\tg, t)$ that has been mapped to $\rglf$, the maximum error $\maxerr_\sasp \triangleq \maxerr(\rglffeas)$ is approximated.

The reachable sets $\hjrset(\tg, t)$ of the true value $V(x,t)$ are shown for different time horizons by the light-colored lines in the bottom row of Figure \ref{fig:SM}; these are solved with DP (via \cite{hjreachpy2021}) over a grid of $100^2$ points in $\ssp$. With the same grid mapped to $\rglf$ in $\ssp$, the reachable sets on $\rglf$ of the SA linear value with antagonistic error $\pj \hjrset_{\sasp, \maxerr}(\tgg^\inn,t)$ are solved with the Hopf formula in parallel (via \cite{hopfreachjl2023}) and also plotted in bottom row of Figure \ref{fig:SM} by the dark-colored lines. As shown, these sets are conservative under-approximations of the true reachable sets. Note, unlike DP, the Hopf formula may solve the value at these points without gridding the entire space of $\sasp$ (or without any grid at all) and in parallel since the value at each point is independent. Solely to elucidate the results, however, on a $100^3$ grid in $\sasp$, the entire reachable set of the SA linear value with antagonistic error $\hjrset_{\sasp, \maxerr}(\tgg^\inn,t)$ is solved with the Hopf formula (in parallel) and plotted in the top row of Figure \ref{fig:SM} with a contour highlighting the intersection of $\hjrset_{\sasp, \maxerr}(\tgg^\inn,t)$ and $\rglf$.

\subsection{Van der Pol System} \label{sec:VdP}
 
To observe the results applied to various lifting functions, consider the Van der Pol system with control,
\begin{eqnarray}
    \dot{x} = \begin{bmatrix} x_1 \\ \mu(1 - x_1^2)x_2 - x_1 \end{bmatrix} + \begin{bmatrix} 0 \\ 1 \end{bmatrix} u
\end{eqnarray}
with $\mu=1$ and $u \in \cset \triangleq \{\vert u \vert \le 1/2 \}$. Let the game in this setting be such that the control aims to drive trajectories away from $\tg \triangleq \{x^\top x \le 1\}$ at time $T$ (\textit{Avoid} game). There is no disturbance in this game i.e. it is an optimal control problem that will, via our method, be converted into a game in the SA space to account for the error of any SA linear model.

Consider state augmentations of this system defined by lifting functions of polynomials of degrees three ($n_k = 10$) and four ($n_k = 15$), and radial basis functions (RBFs) with Gaussian kernels with five ($n_k = 7$) and nine centers ($n_k = 11$). The linear models for the SA systems are generated by a standard EDMD method which uses the $L_2$ error for fitting a linear model to a random trajectory sample of 2000 points (via \cite{pan2023pykoopman}).  For all lifting functions, let $\tgg^\out \triangleq \{ g^\top Q g \le 10/9 \} $ with $Q = \text{Diag}([1,1, \eta \bm{1}])$ with $\bm{1} \in \mathbb{R}^{\dimsa - \dimst}$ and $\eta=10$, defining $(\tgg^\out \cap \rglf) \supset (\tgg \cap \rglf)$. 

\begin{figure}[t]
    \centering
    \includegraphics[width=\linewidth]{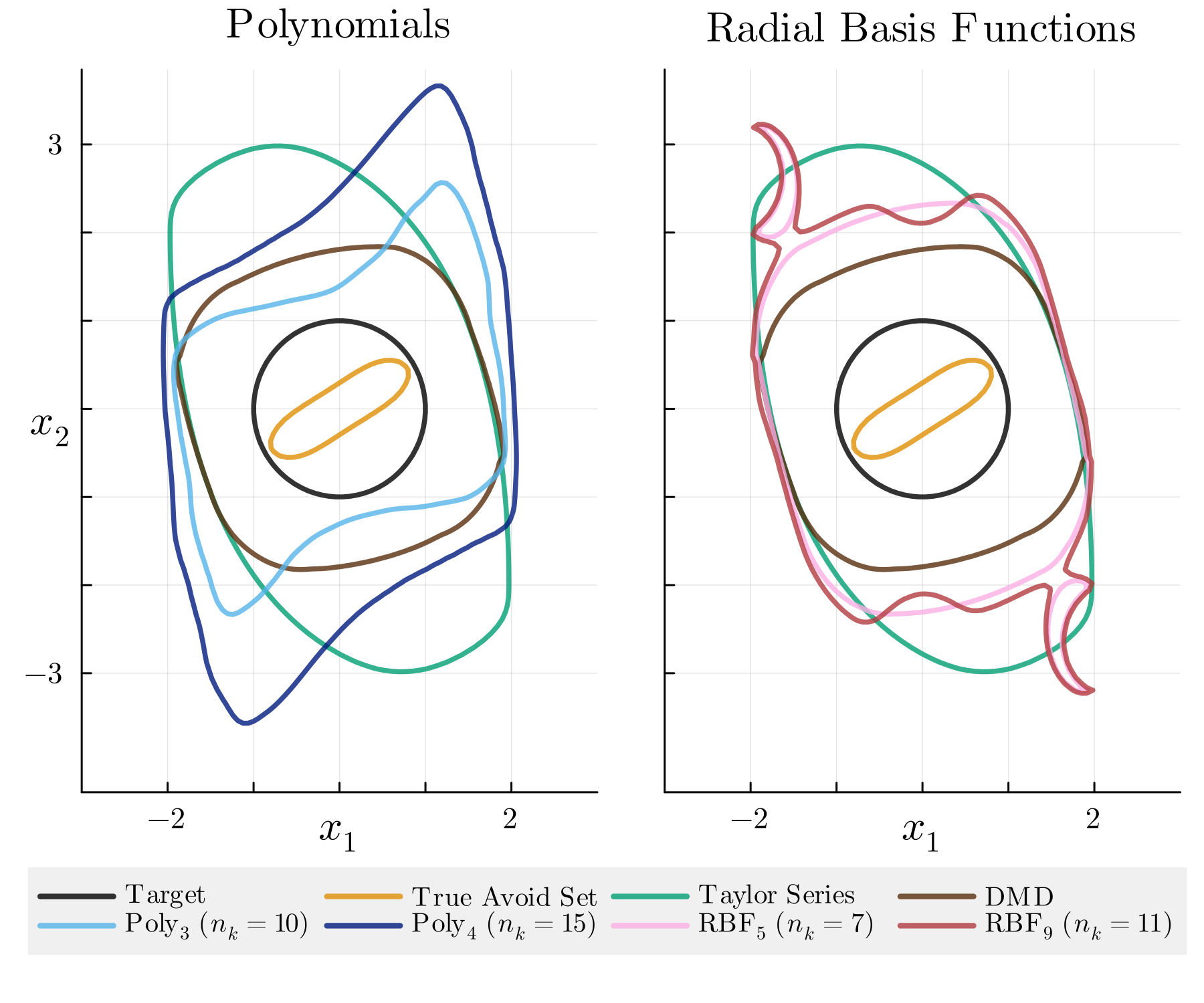}    
    \caption{\textbf{\textit{Avoid} game in the Van der Pol system (Sec.\ref{sec:VdP}) with various lifting functions $\Psi$. 
    }
    In both plots, the target $\tg$ and the (true) backward avoid set $\hjrset^-(\tg,t)$ are shown in black and gold resp. 
    As baselines, the avoid sets of the games with antagonistic error (without lifting) $\hjrset^-_{\maxerr}$ with the Taylor Series (TS) and Dynamic Mode Decomposition (DMD) linearizations are shown in green and brown.
    On the left, the projections restricted to the manifold of the avoid sets of state-augmented linear models with antagonistic error, $\pj \hjrset^-_{\sasp, \maxerr}$, are shown for two polynomial lifting functions of degrees three and four respectively in shades of blue. On the right, the $\pj \hjrset^-_{\sasp, \maxerr}$ are shown for two RBF lifting functions with five and nine centers respectively in shades of red. Theorem~\ref{thm:HJR-SAS} guarantees that for any $\lf$ satisfying the given assumptions, $\pj \hjrset^-_{\sasp, \maxerr} \supset \hjrset^-$, therefore the intersection of all $\pj \hjrset^-_{\sasp, \maxerr}$ is a valid conservative envelope of the avoid set $\hjrset^-(\tg,t)$. Note, the tightness of the over-approximation depends on the maximum error $\maxerr$ for a linear model with any given lifting function.
    }
    \label{fig:VDP}\vspace{-1em}
\end{figure}

In the same manner as in Sec.\ref{sec:SM}, the tube $\ftube(\tg, t)$ is conservatively solved via \cite{JuliaReach19}, the maximum error $\maxerr_\sasp \triangleq \maxerr(\rglffeas)$ is approximated on a grid of $\ftube(\tg, t)$ mapped to $\rglf$, and then $\hjrset_{\sasp, \maxerr}^-$ is solved with the Hopf formula and compared to the DP-based ground truth $\hjrset^-$ at $t=T-1$. In addition, the Taylor series (TS) and dynamic mode decomposition (DMD) non-augmented solutions $\hjrset_{\maxerr}^-$ in $\ssp$ are solved with the Hopf formula and included for comparison. The solutions are shown in Figure ~\ref{fig:VDP}.

Interestingly in this example, while the mean error on the evolution of identity states decreases with higher $n_k$ (not shown), the maximum error does not, and it can be seen that the highest $n_k$ does not give the tightest over-approximation. This is consistent with the \textit{asymptotic} nature of the limit to the Koopman operator \cite{korda2018convergence, nuske2023finite, bramburger2023auxiliary}. Moreover, this is affected by the natural tendency of the L$_2$ metric to scale with increased dimension. This could be improved by SA fitting with the L$_\infty$ metric or the consistency index \cite{haseli2024modeling} but we leave this to future work.

\section{CONCLUSION}\label{sec:conclusion}

In this work, we have devised the construction of a differential game for state-augmented linear models with antagonistic error. Moreover, we prove the corresponding value is conservative with respect to the true value, and by construction, may be used to derive an optimal controller that is guaranteed to succeed in the true dynamics. This is valuable considering that state-augmented systems may have significantly improved accuracy, and the results are amenable to combination (with union or intersection for \textit{Reach} and \textit{Avoid} resp.). Moreover, all of the extensions to further reduce the error in \cite{sharpless2024conservative} are applicable to the current setting and we leave this to future work. Notably, this method also offers a novel way to use linear differential games to approximate solutions in a non-convex fashion. Future work may include extensions to probabilistic error bounds, neural net lifting functions, and non-state inclusive augmented space.



\bibliographystyle{IEEEtran}
\bibliography{main}
\end{document}